\documentclass[conference,a4paper]{IEEEtran}
\IEEEoverridecommandlockouts

\usepackage{cite}
\usepackage{amsmath,amssymb,amsfonts}
\usepackage{algorithmic}
\usepackage{graphicx}
\usepackage{textcomp}
\usepackage{xcolor}
\usepackage{threeparttable}
\usepackage{array}

\def\BibTeX{{\rm B\kern-.05em{\sc i\kern-.025em b}\kern-.08em
    T\kern-.1667em\lower.7ex\hbox{E}\kern-.125emX}}

\IEEEaftertitletext{\vspace{-1\baselineskip}}
\begin{document}

\title{PSA: A Pruning-Friendly Sparse-Aware Accelerator for Efficient Sparse Convolution in SNNs\\

\thanks{This work was supported in part by the National Natural Science Foundation of China under grant 62304203 and in part by the International Campus of ZJU International Research Collaboration Seed Project under grant GJ202509.}
}

\author{\IEEEauthorblockN{Sanhe Fu}
\IEEEauthorblockA{\textit{ZJUI-UIUC Institute,} \\
\textit{Zhejiang University,} \\
Haining, China \\
sanhe1.25@intl.zju.edu.cn}
\and
\IEEEauthorblockN{Kainan Wang}
\IEEEauthorblockA{
\textit{College of Information Science} \\
\textit{and Electronic Engineering,} \\
\textit{Zhejiang University,} \\
Hangzhou, China \\
kainan.22@intl.zju.edu.cn}
\and
\IEEEauthorblockN{Bo Wang}
\IEEEauthorblockA{\textit{Senior Member, IEEE} \\
\textit{Singapore University of} \\
\textit{Technology and Design,} \\
Singapore \\
bo\_wang@sutd.edu.sg}
\and
\IEEEauthorblockN{Aili Wang}
\IEEEauthorblockA{\textit{Member, IEEE} \\
\textit{ZJUI-UIUC Institute,} \\
\textit{Zhejiang University,} \\
Haining, China \\
ailiwang@intl.zju.edu.cn}
}

\setlength{\textfloatsep}{1pt}
\setlength{\floatsep}{1pt}
\setlength{\intextsep}{1pt}
\setlength{\dbltextfloatsep}{1.5pt}
\setlength{\dblfloatsep}{1.5pt}
\setlength{\abovecaptionskip}{1pt}
\setlength{\belowcaptionskip}{1pt}

\maketitle

\begin{abstract}
	Spiking neural networks (SNNs) are promising for energy-efficient artificial intelligence due to their event-driven sparse computation, and neuron pruning can further reduce redundant operations. However, in practical hardware accelerators, spike sparsity and pruning-induced sparsity produce highly irregular sparse workloads that are difficult to execute efficiently. Existing approaches do not jointly support sparse-input execution, low-overhead sparse convolution lowering and pruning-friendly execution. This paper presents PSA, a Pruning-friendly Sparse-aware Accelerator for efficient sparse convolution in SNNs. PSA features a Sparse-Aware Merge-Accumulation (SAMA) dataflow that aligns sparse spike indices and converts irregular activations into continuous computation streams for efficient sparse convolution execution. It further introduces a Sparse Implicit Group-wise im2col (SIG-im2col) convolution lowering mechanism that converts sparse convolution into SpMM on-chip with low mapping cost and memory overhead. In addition, PSA integrates a lightweight pruning-friendly gating mechanism to directly support pruned SNNs. Results on ResNet-18/CIFAR-10 show that PSA achieves a 4.45x improvement in energy efficiency and an 18.52x reduction in energy-delay product (EDP) over the dense baseline.
\end{abstract}


\section{Introduction}

Spiking neural networks (SNNs) are promising for edge intelligence because their event-driven computation naturally produces sparse activity, offering the potential for high energy efficiency. To exploit these advantages, recent hardware accelerators have been developed to improve spike processing efficiency and reduce energy consumption \cite{xu2025spikex,fang2024cicc,wang2025tcasi}.

However, efficiently executing sparse convolutions in SNN accelerators remains challenging. Sparse spike activity produces highly irregular nonzero activation patterns, which often lead to low processing-element (PE) utilization and inefficient memory access on conventional datapaths. A natural way to regularize convolution is to lower it into GEMM-style computation. However, conventional im2col is typically designed for dense workloads and requires materializing the unfolded intermediate feature map, causing data expansion and additional memory and scheduling overhead. In addition, neuron pruning, often used to further compress SNN models, introduces extra sparsity and irregularity into the computation graph. As a result, although many operations in SNN workloads can theoretically be skipped, translating spike and pruning sparsity into practical throughput and energy efficiency gains remains difficult.

Recent studies have explored several directions to improve efficiency. Designs such as SpikeX exploit sparsity-aware spike scheduling \cite{xu2025spikex}, Fang \textit{et al.} propose a specialized compute array for sparse workloads \cite{fang2024cicc}, and GAMMA uses Gustavson's algorithm to achieve high throughput \cite{zhang2021gamma}. Another line of work accelerates convolution lowering: SPOTS maps sparse convolutions to SpMM and executes them on a systolic GEMM array \cite{soltaniyeh2022spots}, while Fornt \textit{et al.} perform on-the-fly im2col to reduce data movement \cite{fornt2023IM2COL}. Meanwhile, neuron pruning techniques have also been explored to reduce redundant computation in SNNs \cite{shi2024pruning,li2024activitypruning,belhadi2025lightprune}. However, existing works optimize spike sparsity exploitation, convolution lowering, and neuron pruning in separate ways. How to support them together in a single low-overhead hardware execution path remains an open challenge.

To address this gap, we propose PSA, a Pruning-friendly Sparse-aware Accelerator for efficient sparse convolution in SNNs. PSA introduces a Sparse-Aware Merge-Accumulation (SAMA) dataflow that aligns and aggregates sparse spike indices from multiple input channels to form continuous computation streams, improving PE utilization under irregular activations. It further incorporates a Sparse Implicit Group-wise im2col (SIG-im2col) convolution lowering mechanism that performs hardware-resident sparse-convolution-to-SpMM mapping in a group-wise manner without materializing the intermediate feature map, thereby reducing mapping cost and memory overhead. In addition, PSA integrates an on-chip pruning-friendly gating mechanism that represents pruning results as lightweight index streams and filters invalid output updates during execution. The main contributions of this work are as follows: 1) a sparse-aware merge-accumulation dataflow for efficient streaming execution of irregular sparse convolutions; 2) a sparse implicit group-wise im2col lowering method with low mapping cost and memory overhead; and 3) a pruning-friendly execution mechanism that directly supports neuron-pruned SNN deployment with low hardware overhead.

\begin{figure*}[t]
	\centering
	\includegraphics[width=1\linewidth]{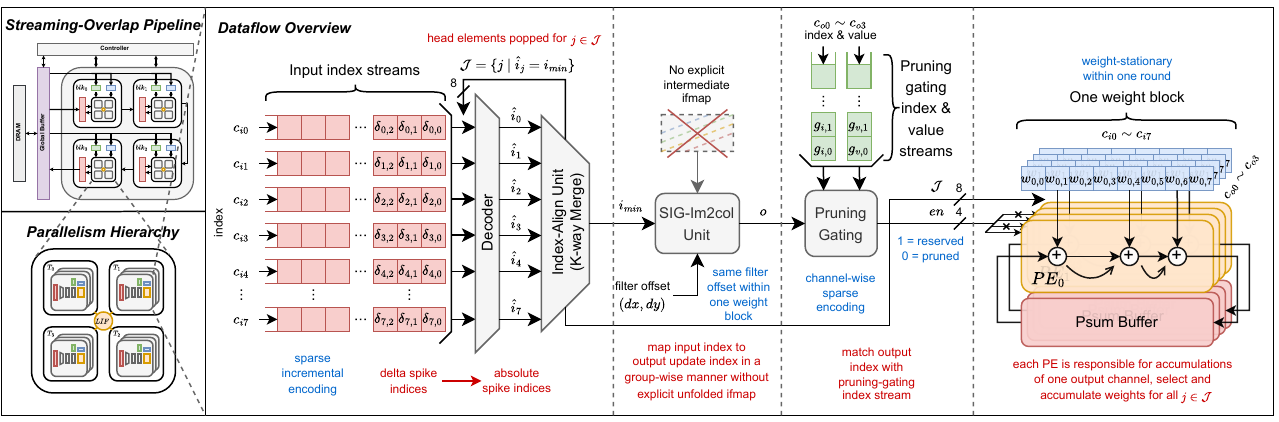}
	\caption{System overview of PSA: SAMA dataflow, SIG-im2col lowering, and pruning-friendly execution.}
	\label{fig:system-overview}
\end{figure*}

\section{Dataflow and Microarchitecture}

\subsection{System Overview and Data Organization of PSA}
Fig.~\ref{fig:system-overview} shows the execution flow of PSA. PSA integrates three tightly coupled mechanisms into a unified sparse-convolution pipeline, allowing sparse-input alignment, sparse convolution lowering, and pruning-friendly filtering to be performed in one hardware path. Spike activations are stored as relative offsets between consecutive nonzeros and reconstructed into absolute indices only when required for alignment, while pruning results are represented as channel-wise sparse streams with shared indices and binary gating values. In this way, PSA processes both spike sparsity and pruning sparsity as indexed streams without dense feature expansion.

PSA organizes computation around fixed-granularity weight blocks in a weight-stationary manner. In each round, one weight block is stored in local buffers and reused while the corresponding activation and gating streams are fetched on demand. Under a typical configuration, one weight block forms an 8$\times$4 channel tile with a fixed filter offset. Input indices are first aligned by SAMA, then mapped to output-update indices by SIG-im2col, and finally filtered by pruning gating before accumulation in the PE array. Only valid output updates are forwarded to compute units for psum generation.

\begin{figure*}[t]
	\centering
	\includegraphics[width=1\linewidth]{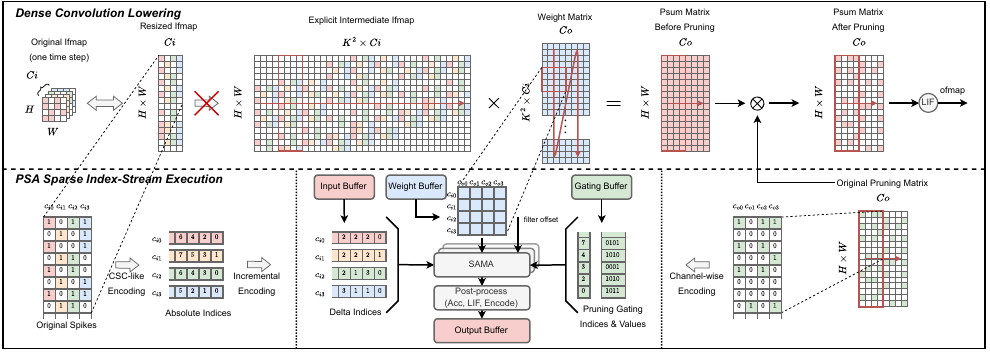}
	\caption{Data organization and tiling correspondence between conventional dense convolution lowering and PSA sparse index-stream execution.}
	\label{fig:data-organization-tiling-overview}
\end{figure*}

Fig.~\ref{fig:data-organization-tiling-overview} further illustrates the correspondence between conventional dense im2col convolution lowering and PSA sparse index-stream execution. Unlike the conventional flow that explicitly materializes an intermediate feature map, PSA directly operates on incrementally encoded spike streams from the original feature map and performs SIG-im2col only after on-chip index alignment. As a result, tiling is defined over sparse index streams, weight blocks, and pruning-friendly output updates rather than over dense unfolded feature tiles.

\begin{figure}[b]
	\centering
	\includegraphics[width=1\linewidth]{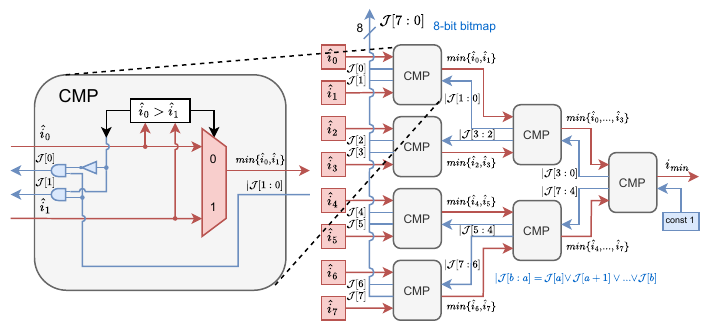}
	\caption{Hardware structure of the Index-Align unit for selecting $i_{\min}$ and aggregation set $\mathcal{J}$.}
	\label{fig:index-align-unit}
\end{figure}

\subsection{Index-Align via K-way Merge Mechanism}

This subsection presents SAMA's k-way merge mechanism for aligning sparse indices across different input columns. For each selected input column $j\in[0,k-1]$, the activation spikes are represented as a linear index sequence:
\begin{equation}
	\mathcal{I}_j=\{i_{j,0}, i_{j,1}, \dots\},
\end{equation}
where $\mathcal{I}_j$ is a strictly increasing input-index stream decoded from the sparse activation encoding.

The k-way merge mechanism leverages the sorted property of each input stream to perform lightweight online alignment. Hardware maintains only the current head input index $\hat{i}_j$ for each input column, and a comparator tree selects the global minimum:
\begin{equation}
	i_{\min} = \min_{j} \hat{i}_j
\end{equation}
All input columns satisfying $\hat{i}_j=i_{\min}$ are grouped into an aggregation set:
\begin{equation}
	\mathcal{J}=\{j \mid \hat{i}_j=i_{\min}\},
\end{equation}
and the pair $(i_{\min}, \mathcal{J})$ is emitted. In hardware, the aggregation set $\mathcal{J}$ is encoded as a k-bit bitmap, where bit $j$ is asserted if  $\hat{i}_j=i_{\min}$. As shown in Fig.~\ref{fig:index-align-unit}, the comparator (CMP) tree generates both the minimum index $i_{\min}$ and the aggregation set $\mathcal{J}$ for tied input streams.  For columns in $\mathcal{J}$, the head element is popped and the pointer advances; other columns remain unchanged. Iterating this procedure produces an aggregated stream in increasing input-index order and aligns events with the same input index (the same pixel position) across different input columns (different input channels).

Notably, the merge operates directly on index streams from the original input feature map (ifmap), rather than on indices from an explicit intermediate ifmap unfolded by traditional im2col. This design enables multiple participating input columns to share the same emitted input index, which is then used by the following SIG-im2col stage to perform group-wise mapping.

\subsection{Sparse Implicit Group-wise Im2col Mapping}
To reduce the lowering overhead from sparse convolution to SpMM, SAMA adopts an on-chip sparse implicit group-wise im2col (SIG-im2col) mapping. Instead of explicitly constructing the unfolded ifmap or relying on off-chip scheduling, SAMA directly maps sparse input indices to output-update indices online and immediately drives the subsequent accumulation.

The mapping is implemented as a lightweight stage after the index-align unit. For each merged aggregation group $(i_{\min}, \mathcal{J})$ generated by k-way merging, SAMA performs the SIG-im2col mapping only once on the shared input index $i_{\min}$ to obtain the output-update index $o$ and a validity flag, and then reuses $o$ for all participating input columns in $\mathcal{J}$. With a fixed filter offset $(d_x,d_y)$, valid output-update indices are generated in increasing order as input indices increase. Therefore, SAMA can perform k-way merge directly in the original input-index domain, rather than in the index domain of an explicitly unfolded ifmap.

If the mapping is invalid, e.g., due to boundary violation or stride divisibility failure, the group is skipped. Otherwise, $o$ is used as the destination psum index, and one accumulation is triggered for each input column in $\mathcal{J}$.

The mapping is defined as follows. Let the pre-padding input feature map have spatial size $W\times H$, and let a sparse activation be represented by a row-major linear index $i$. Its input coordinate $(x_i,y_i)$ is
\begin{equation}
	x_i = i \bmod W,\quad
	y_i = \left\lfloor \frac{i}{W} \right\rfloor .
\end{equation}
For a fixed kernel position with offset $(d_x,d_y)$ and stride $s$, the output coordinate $(x_o,y_o)$ is
\begin{equation}
	x_o = \frac{x_i - d_x}{s},\quad
	y_o = \frac{y_i - d_y}{s}.
\end{equation}
The mapping is valid only if
\begin{equation}
	(x_i - d_x)\bmod s = 0,\quad
	(y_i - d_y)\bmod s = 0,
\end{equation}
\begin{equation}
	0 \le x_o < W_o,\quad
	0 \le y_o < H_o,
\end{equation}
where $(W_o,H_o)$ is the output feature-map size. For a valid mapping, the output linear index is
\begin{equation}
	o = y_o \cdot W_o + x_o.
\end{equation}

Compared with traditional im2col, which maps and buffers the entire unfolded ifmap, the proposed SIG-im2col method maps only the shared input index once per emitted aggregation group (group-wise) and eliminates the need to store any intermediate ifmap. Therefore, the mapping cost scales with the number of emitted aggregation groups rather than the size of the unfolded ifmap.

\subsection{Pruning-Deployment-Friendly On-Chip Filtering Mechanism}

To support pruning-friendly deployment with low overhead, SAMA adopts an on-chip filtering mechanism based on streaming gating indices and values. We logically model unstructured neuron pruning as a binary gating matrix, and store it in a channel-wise sparse indexed format. Each gating index denotes an output spatial position, while its associated gating value indicates which output channels in the current block remain active after pruning.

In SAMA, the gating stream constrains the active output set during accumulation. After an output-update index $o$ is generated, the gating unit matches $o$ against the current gating window. If $o$ is not covered by the active gating entries, the update at $o$ is skipped for the current output-channel block. Otherwise, only those output channels with a gating value 1 are enabled, and the remaining channels are skipped.

\begin{figure}
	\centering
	\includegraphics[width=1\linewidth]{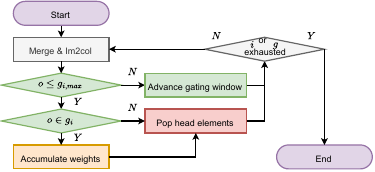}
	\caption{Control flow of the pruning-friendly on-chip filtering mechanism.}
	\label{fig:gating-process}
\end{figure}

To avoid scanning all pruning results, SAMA maintains only a small on-chip window over the strictly increasing gating-index stream. Since the emitted output-update indices are generated in increasing order, the gating window only needs forward advancement and never rolls back. When $o$ exceeds the current window upper bound ($o > g_{i,max}$), the gating unit advances the window on demand to cover subsequent output-index space while keeping the head elements unchanged, otherwise, each input column in $\mathcal{J}$ pops its head element and advances its pointer, as illustrated in Fig.~\ref{fig:gating-process}.

By matching the output-update stream in the output-index domain within a small on-chip window, the proposed mechanism avoids dense mask storage and full-space scanning for pruning-friendly inference. As a result, pruning results can be efficiently mapped to on-chip execution with low control and memory overhead.

\section{Implementation Results}

\begin{figure}[b]
	\centering
	\includegraphics[width=1\linewidth]{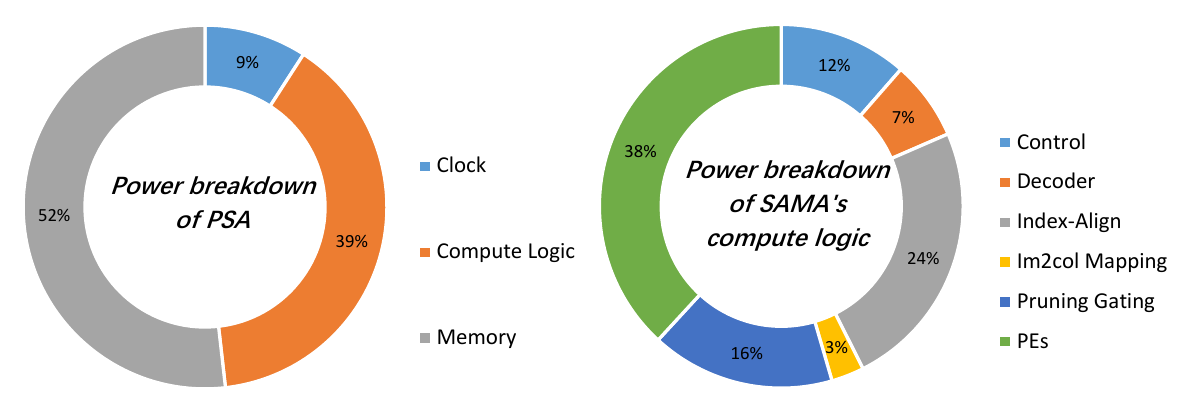}
	\caption{Power breakdown of PSA and internal SAMA compute logic.}
	\label{fig:pwr-breakdown}
\end{figure}

\begin{figure}
	\centering
	\includegraphics[width=1\linewidth]{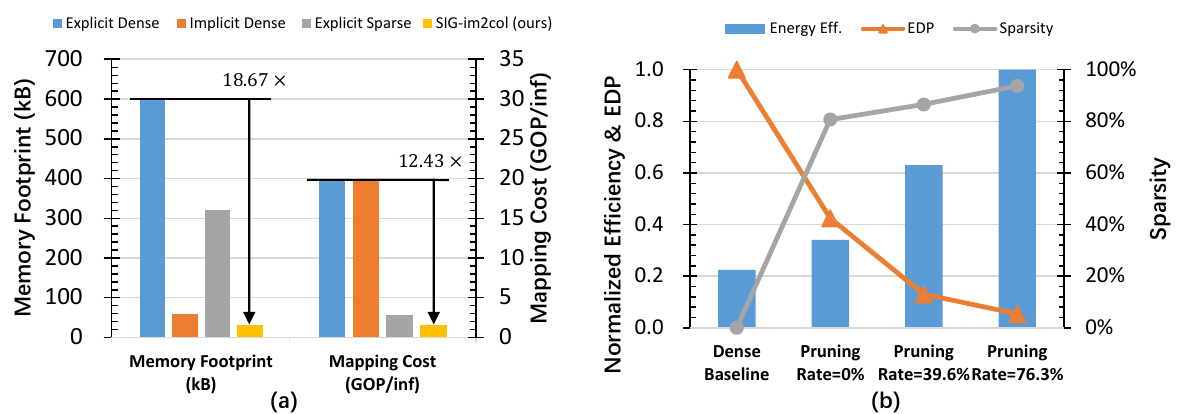}
	\caption{(a) Comparison of ifmap memory footprint (kB) and im2col mapping cost ($\#$mapping OP per inference) between explicit and implicit im2col under dense and sparse settings ($T=4$, sparsity $=86.48\%$). (b) Comparison of energy efficiency and EDP (normalized) under different pruning rates.}
	\label{fig:im2col-pruning-results}
\end{figure}

We synthesized PSA in a 28\,nm CMOS technology at 500\,MHz and 0.72\,V, and the overall implementation results are summarized in Table~\ref{tab:big-table}. Because PSA is designed to jointly support sparse-input execution, sparse convolution lowering, and pruning-friendly deployment, we evaluate it from three complementary aspects: the implementation overhead of the sparse-control path, the cost reduction enabled by SIG-im2col lowering, and the resulting system-level efficiency gains under pruning-friendly execution.

Fig.~\ref{fig:pwr-breakdown} shows the power breakdown of PSA. The sparse-control path, including the decoder, index-align unit, SIG-im2col mapper, and pruning-gating logic, accounts for a limited portion of total power compared with the PE array and memory, indicating that the proposed sparse-execution support does not impose excessive control overhead. The internal breakdown of SAMA further shows that the added control and mapping logic remains well contained within the overall compute cost.

Fig.~\ref{fig:im2col-pruning-results}(a) evaluates the proposed SIG-im2col mechanism. By avoiding explicit materialization of the unfolded intermediate feature map and performing group-wise online mapping directly on sparse index streams, PSA reduces the ifmap memory footprint by 18.67$\times$ and the mapping cost by 12.43$\times$ compared with a dense explicit im2col baseline. These results demonstrate that the proposed lowering scheme significantly reduces intermediate storage and mapping overhead for convolution lowering.

Fig.~\ref{fig:im2col-pruning-results}(b) evaluates pruning-friendly deployment efficiency. As the pruning rate increases, the normalized energy efficiency improves consistently, while the energy-delay product (EDP) decreases accordingly. Compared with the dense baseline, PSA achieves 4.45$\times$ higher energy efficiency and 18.52$\times$ lower EDP at the highest evaluated pruning setting. This trend indicates that PSA can effectively translate pruning-induced sparsity into practical hardware efficiency gains with limited additional deployment overhead.

\begin{table}[h]
	\centering
	\begin{threeparttable}
		\caption{Performance Summary and Comparison With Prior Work}
		\label{tab:big-table}
		\setlength{\tabcolsep}{0pt}
		\renewcommand{\arraystretch}{1}
		\scriptsize
		\begin{tabular}{
				|>{\centering\arraybackslash}m{2.2cm}| >{\centering\arraybackslash}m{1.3cm}| >{\centering\arraybackslash}m{1.3cm}| >{\centering\arraybackslash}m{1.3cm}| >{\centering\arraybackslash}m{1.3cm}| >{\centering\arraybackslash}m{1.3cm}|
			}
			
			\hline & \textbf{THIS WORK} & \textbf{HPCA'25\cite{wei2025prosperity}} & \textbf{TVLSI'24\cite{xie2024tvlsi}} & \textbf{ISSCC'23\cite{kim2023isscc}}& \textbf{JETCAS'19\cite{chen2019jetcas}} \\
			\hline \textbf{Technology [nm]} & 28 & 28 & 28 & 28 & 65 \\ \hline \textbf{Area [mm$^2$]} & 2.55 & 0.53 & N/A & 20.25 & N/A \\
			\hline \textbf{Supply Voltage [V]} & 0.72 & N/A & 0.90 & 0.7 & N/A \\
			\hline \textbf{Frequency [MHz]} & 500 & 500 & 650 & 50 & 200 \\
			\hline \textbf{PEs} & 240 & 128 & N/A & N/A & 192 \\ \hline \textbf{Type} & \textbf{SNN} & SNN & SNN & SNN & CNN \\
			\hline \textbf{Sparsity Support} & \textbf{Input + Neuron}\tnote{b} & Input & Input & Input & Activation\tnote{d} + Weight \\
			\hline \textbf{Model \& Dataset} & \textbf{ResNet-18: CIFAR-10} & VGG-16: CIFAR-100 & ResNet-18: CIFAR-10 & ResNet-50: ImageNet & AlexNet: ImageNet \\
			\hline \textbf{Peak Performance}\tnote{a} \textbf{[GOPS]} & \textbf{843.93} & 390.10 & 746.00 & N/A & 153.60 \\
			\hline \textbf{Energy Efficiency [TOPS/W]} & \textbf{24.09}\tnote{c} & 0.30 & 36.89 & 20.80 & 2.5\tnote{e} \\
			\hline \textbf{Area Efficiency [GOPS/mm$^2$]} & \textbf{331.12}\tnote{c} & 737.43 & N/A & N/A & N/A \\
			\hline
			
		\end{tabular}
		
		\begin{tablenotes}
			\tiny \item[a] OP definitions may differ across works; GOPS and TOPS/W are reported according to each paper's original convention. In this work, \#OP = \#(dense ACC per time step) $\times$ \#(time step).
			\item[b] Support for neuron pruning (neurons are sparse after pruning) of SNN. 
			\item[c] Energy Eff. \& Area Eff. @ Pruning Rate = 76.3\%, Input Sparsity = 93.67\%.
			\item[d] Support for sparse activation processing in activation layer of ANN. 
			\item[e] For fairness, energy efficiency is normalized across technology nodes following the estimation methodology used in prior accelerator works.
		\end{tablenotes}
	\end{threeparttable}
\end{table}

\section{Conclusion}
SNNs offer promising energy efficiency due to their event-driven nature, and neuron pruning can further unlock this efficiency potential. However, these sparsity benefits are difficult to realize on real hardware. This paper presented PSA, a pruning-friendly sparse-aware accelerator for efficient sparse convolution in SNNs. PSA combines the SAMA dataflow, SIG-im2col convolution lowering, and pruning-friendly on-chip filtering to jointly support low-overhead sparse execution and pruning deployment. Implemented in TSMC 28\,nm, PSA achieves 843.93\,GOPS and 24.09\,TOPS/W, while improving energy efficiency by 4.45$\times$ and reducing EDP by 18.52$\times$ over the dense baseline. These results demonstrate that spike and pruning sparsity can be effectively translated into hardware efficiency gains through the proposed architecture.

\bibliographystyle{IEEEtran}
\bibliography{ref}

\end{document}